\documentclass{elsart}
\usepackage{graphicx}
\usepackage{dcolumn}
\usepackage{bm}
\usepackage{amsmath}
\usepackage{amsfonts}
\usepackage{amssymb}

\begin{document}
\begin{frontmatter}
\title{Sub-ballistic behavior in the quantum kicked rotor}
\author{A. Romanelli\thanksref{alejo}},
\author{A. Auyuanet\thanksref{adriana}},
\author{R. Siri\thanksref{pollo}},
\author{V. Micenmacher\thanksref{CERP}}
\address{Instituto de F\'{\i}sica, Facultad de Ingenier\'{\i}a\\
Universidad de la Rep\'ublica\\ C.C. 30, C.P. 11000, Montevideo,
Uruguay}
\thanks[alejo]{email: alejo@fing.edu.uy}
\thanks[adriana]{email: auyuanet@fing.edu.uy}
\thanks[pollo]{email: rsiri@fing.edu.uy}
\thanks[CERP]{Permanent address: Centro Regional de Profesores del Este,
Maldonado, \\Uruguay, email: vmd@fing.edu.uy}
\date{\today}
\begin{abstract}
We study the resonances of the quantum kicked rotor subjected to an
excitation that follows an aperiodic Fibonacci prescription. In such
a case the secondary resonances show a sub-ballistic behavior like
the quantum walk with the  same aperiodic prescription for the coin.
The principal resonances maintain the well-known ballistic behavior.
\end{abstract}
\begin{keyword}
Kicked rotor; Quantum walk
\end{keyword}
\end{frontmatter}


The quantum kicked rotor (QKR) is considered as the paradigm of
periodically driven systems in the study of chaos at the quantum
level \cite{CCI79}. Two important characteristics of the behavior of
the quantum kicked rotor are dynamical localization (DL) and the
ballistic spreading of the resonances \cite{Izrailev}. These
behaviors are quite different and depend on whether the period of
the kick $\tau$ is a rational or irrational multiple of $4\pi$ (in
convenient units). For rational multiples, the behavior of the
system is resonant and has no classical analog. For irrational
multiples the average energy of the system grows in a diffusive
manner for a short time and then the diffusion stops and
localization appears. From the theoretical point of view the two
types of values of $\tau$ determine the spectral properties of the
Hamiltonian, for irrational multiples the energy spectrum is purely
discrete and for rational multiples it contains a continuous part.
Both quantum resonance and DL can be seen as interference phenomena,
the first is a constructive interference effect and the second is a
destructive one.

We recently developed \cite{generali} a new generalized discrete
time quantum walk (QW) on the line and showed that this model has a
dynamical behavior analogous to that of a QKR: depending on the
values of a parameter there are either quantum resonances or DL.
This modified QW has been mapped into a one-dimensional Anderson
model \cite{Anderson}, as was previously done in the case of the QKR
\cite{Grempel}. For some exceptional values of the parameter, which
correspond to resonant behavior, the standard deviation grows
linearly in time. We showed that the usual discrete time QW on the
line becomes a particular case of resonance of the modified QW.

The concept of QW on the line is a subject that has drawn much
attention in quantum computing \cite{Kempe}. It has been introduced
in 1993 \cite{Aharonov,Godoy}, as counterpart of the classical
random walk. A classical random walk is defined in terms of the
probabilities for a particle to do a step to the left or to the
right but the QW is described in terms of probability amplitudes.
Many classical algorithms are based on classical random walks, then
it is possible that future quantum algorithms will be based on the
quantum random walk. One of the most striking properties of the QW
is its ability to spread over the line linearly in time as
characterized by the standard deviation $\sigma(t)\sim t$, while its
classical analog spreads out as the square root of time
($\sigma(t)\sim t^{1/2}$).

In this work we establish another aspect of the parallelism between
the QKR and the QW following the proposal of Ribeiro $et$ $al$.
\cite{Ribeiro}. In \cite{Ribeiro} the quantum coin operator of the
QW is arranged in aperiodic sequences using the Fibonacci
prescription and this leads to a sub-ballistic wave function
spreading ($\sigma(t)\sim t^{c},1/2<c<1$).  We show that the
secondary resonances of the QKR, excited with the same Fibonacci
prescription, have also sub-ballistic behavior like the usual QW,
but the primary resonances maintain the well-known ballistic
behavior. Then the parallelism with the usual QW on the line is
restricted to the secondary resonances of the QKR. On the other hand
Casati $et$ $al$. \cite{Casati} studied the dynamics of the QKR also
kicked according to a Fibonacci sequence, but outside the resonant
regime, and they found sub-diffusive behavior for small kicking
strengths; in this sense our work can be considered as complementary
to theirs.

The QKR Hamiltonian is
\begin{equation}
H=\frac{P^{2}}{2I}+K\cos\theta\sum_{n=1}^{\infty}\delta(t-nT) \label{qkr_ham}%
\end{equation}
where the external kicks occur at times $t=nT$ with $n$ integer and
$T$ the kick period, $I$ is the moment of inertia of the rotor, $P$
the angular momentum operator, $K$ the strength parameter and
$\theta$ the angular position. In the angular momentum
representation, $P|\ell\rangle=\ell \hbar|\ell\rangle$, the
wave-vector is $|\Psi(t)\rangle=\sum_{\ell=-\infty
}^{\infty}a_{\ell}(t)|\ell\rangle$ and the average energy is
$E(t)=\left\langle \Psi\right|  H\left|  \Psi\right\rangle
=\varepsilon \sum_{\ell=-\infty}^{\infty}\ell^{2}\left|
a_{\ell}(t)\right|  ^{2}$, where $\varepsilon=\hbar^{2}/2I$. Using
the Schr\"{o}dinger equation the quantum map is readily obtained
from the Hamiltonian (\ref{qkr_ham})
\begin{equation}
a_{\ell}(t_{n+1})=\sum_{j=-\infty}^{\infty}U_{\ell j}a_{j}(t_{n}) \label{mapa}%
\end{equation}
where the matrix element of the time step evolution operator $U(\kappa)$ is
\begin{equation}
U_{\ell j}=i^{-(j-\ell)}e^{-ij^{2}\varepsilon T/\hbar}\,J_{j-\ell}(\kappa),
\label{evolu}%
\end{equation}
$J_{m}$ is the $m$th order cylindrical Bessel function and the
argument is the dimensionless kick strength $\kappa\equiv K/\hbar$.
The resonance condition does not depend on $\kappa$ and takes place
when the frequency of the driving force is commensurable with the
frequencies of the free rotor. Inspection of eq.(\ref{evolu}) shows
that the resonant values of the scale parameter
$\tau\equiv\varepsilon T/2\hbar$ are the set of the rational
multiples of 4$\pi$, $\tau=4\pi$ $p/q$. In what follows we assume,
that the resonance condition is satisfied, therefore the evolution
operator depends on $\kappa$, $p$ and $q$. We call a resonance
primary when $p/q$ is an integer and secondary when it is not.

\begin{figure}[tbh]
\begin{center}
\includegraphics[scale=0.33]{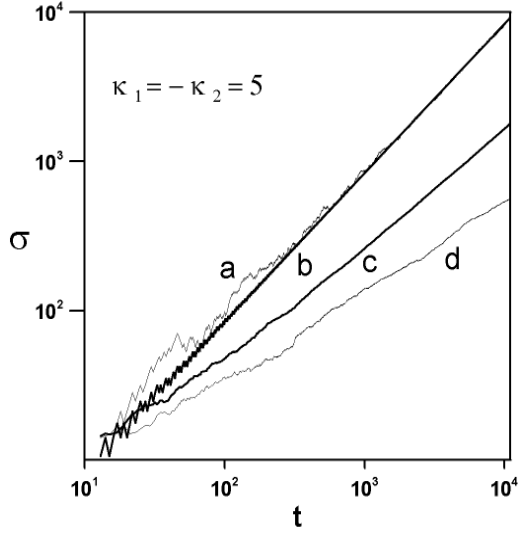}
\end{center}
\caption{Standard deviation $\sigma(t)$ as a function of time. a) random
sequence for the primary resonance $p/q=1$; b) periodical sequence for the
primary resonance $p/q=1$; c) Fibonacci sequence for the secondary resonance
$p/q=1/3$; d) random sequence for the secondary resonance $p/q=1/3$. }%
\label{fig:sigma}%
\end{figure}With the aim to generate the dynamics of the system we consider
two values of the strength parameter $\kappa$, $\kappa_{1}$ and
$\kappa_{2}$ and combine the corresponding time step operators
$U\left(\kappa_{1}\right)$ and $U\left(\kappa_{2}\right)$ in a large
sequence that takes three different forms, namely periodic, random
and quasi-periodic. In this way we generate three types of unitary
evolution operators. With these operators we compute, for several
thousands of $T$, the wave function spreading as measured by the
exponent $c$ in $\sigma(t)=\sqrt{\sum_{\ell=-\infty}^{\infty}\ell
^{2}\left| a_{\ell}(t)\right|  ^{2}}\sim t^{c}$.

\begin{figure}[th]
\begin{center}
\includegraphics[scale=0.38]{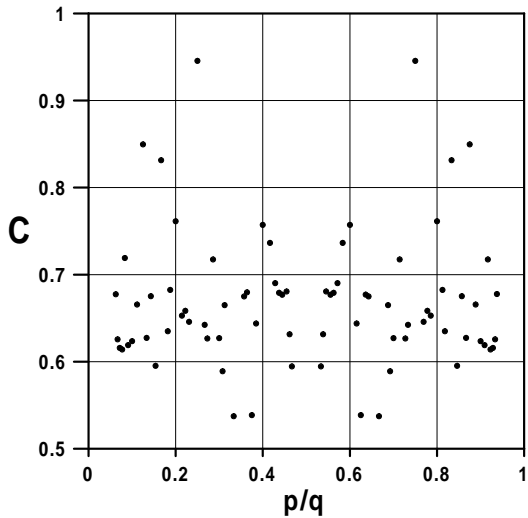}
\end{center}
\caption{The exponent $c$ as a function of the ratio $p/q$ that
identifies the secondary resonance. The probability distribution was
calculated, for the Fibonacci quasi-periodic sequence, with
$\kappa_{1}=5$, $\kappa_{2}=10.$ The value $p/q=1/2$ does not appear
because it corresponds to an antiresonance.
The symmetry of the figure is explained in the text.}%
\label{fig:cp/q}%
\end{figure}
We make a numerical study of the long time behavior of the parameter
$c$ for the three types of sequences. In the periodic case, for
different sequences, the ballistic behavior, $c\sim1$, is still
found for all resonance conditions. In the random case we performed
several calculation where the number of operators
$U\left(\kappa_{1}\right)$ and $U\left(\kappa_{2}\right)$ in the
random sequence were taken in different ratios, obtaining in always
$c\sim1$ for the primary resonances and $c\sim1/2$ for the secondary
ones. The quasi-periodic case was performed using the Fibonacci
prescription given in \cite{Ribeiro} where $U_{n}=U_{n-1}U_{n-2}$
with $U_{0}=U\left(\kappa_{1}\right)$ and
$U_{1}=U\left(\kappa_{2}\right)$;
for example, applying the above rule six times we get $U_{1}U_{0}U_{1}%
U_{1}U_{0}U_{1}U_{0}U_{1}$ and this sequence gives the dynamical
evolution up to $t=8T$. In this case, two types of results have been
obtained: ballistic behavior ($c\sim1$) in the primary resonances
and sub-ballistic behavior ($1/2<c<1$) in the secondary resonances.
The standard deviation is plotted in Fig.~\ref{fig:sigma}, which
displays the qualitative differences between the periodic, random
and quasi-periodic cases.

It is possible to prove analytically why the ballistic behavior is
maintained in the primary resonances for all three types of
sequences. For the primary resonance with the initial condition
$a_{l}\left(  0\right)  =\delta_{l0}$ the solution of the map
eq.(\ref{mapa}), with a given sequence of $U\left( \kappa\right)  $
operators, is $a_{\ell}(t_{n})=\left(  -i\right)  ^{\ell
}\,J_{\ell}\left(  \sum\limits_{j=0}^{n}\kappa(j)\right)  =\left(
-i\right)
^{\ell}\,J_{\ell}\left(  m_{1}\text{ }\kappa_{1}+\text{ }m_{2}\text{ }%
\kappa_{2}\right)  $, where $\kappa(j)$ takes the values
$\kappa_{1}$ or $\kappa_{2}$ involved in the sequence and $m_{1}$
and $m_{2}$ are the numbers of occurrences of $\kappa_{1}$ and
$\kappa_{2}$ respectively, $m_{1}+$ $m_{2}=n$. With the above
amplitudes, the standard deviation is obtained
easily as%
\begin{equation}
\sigma(nT)=\frac{\left(  \alpha\text{ }\kappa_{1}+\text{ }\beta\text{ }%
\kappa_{2}\right)  }{\sqrt{2}}n \label{variance}%
\end{equation}
where $\alpha=m_{1}/n$ and $\beta=m_{2}/n$ are the relative weights
of kicks $\kappa_{1}$ and $\kappa_{2}$ respectively. Thus $c=1$ and,
additionally, we have obtained explicitly the ballistic diffusion
coefficient $D\equiv\sigma(nT)/n=\left( \alpha\kappa
_{1}+\beta\kappa_{2}\right)  /\sqrt{2}$.

The QKR has an exceptional behavior when $p/q=1/2$, called
antiresonance, its characteristic being that the system returns to
the initial state every two periods. In this case, given a sequence
of step operators, the solution of the map eq.(\ref{mapa}), is
$a_{\ell}(t_{n})=\left(  -i\right)  ^{\ell}\,J_{\ell }\left(
\sum\limits_{j=0}^{n}\left(  -i\right)  ^{j}\kappa(j)\right)  $ and
the standard deviation is $\sigma(nT)=\sum\limits_{j=0}^{n}\left(
-i\right) ^{j}\kappa(j)/\sqrt{2}$. Therefore, for all sequences, the
antiresonance in the case $\kappa_{1}\neq$ $\kappa_{2}$ has a
similar behavior as the primary resonances. It is important to
emphasize, that the conceptual key for the ballistic behavior of the
primary resonances (and the antiresonance too) resides in the
commutativity between operators $U\left(  \kappa_{1}\right)  $ and
$U\left(  \kappa_{2}\right)  $. This is not the case for the
secondary resonances where the commutator does not vanish.

\begin{figure}[th]
\begin{center}
\includegraphics[scale=0.38]{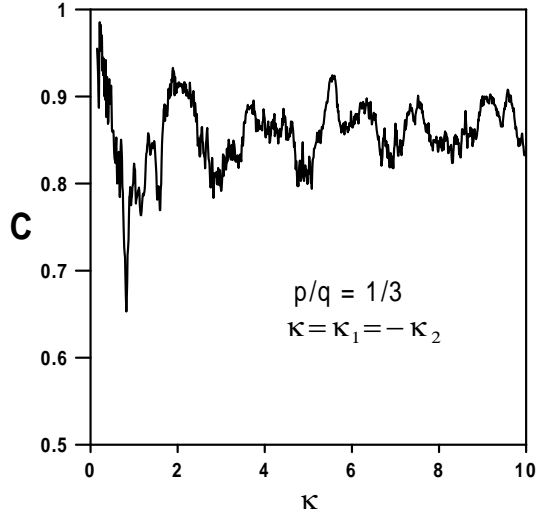}
\end{center}
\caption{The exponent $c$, that characterizes the power law of the
standard deviation, as a function of the strength parameter $\kappa$
of the kicked rotator. The dynamical evolution is obtained by the
operators $U(\kappa_{1})$
and $U(\kappa_{2})$ following a Fibonacci succession.}%
\label{fig:ck1k2}%
\end{figure}

The sub-ballistic behavior takes place for all secondary resonances
in the quasi-periodic case. In Fig.~\ref{fig:cp/q} the exponent $c$
is plotted as a function of the ratio $p/q$ that identifies the
secondary resonance. From this figure we can conclude that the
exponent $c$ depends on both $p$ and $q$, but there is a trivial
symmetry in the time step evolution operator eq.(\ref{evolu}) when
$p/q$ is changed by $(q-p)/q$, this is the reason why $e.g.$ the
value of $c$ for $p/q=1/5$ is the same as for $p/q=4/5$. We have
also found that the dependence of $c$ on the values of $\kappa_{1}$
and $\kappa_{2}$ is not smooth in general. The condition
$\kappa_{1}=\kappa_{2}$ corresponds to the periodic case and the
expected ballistic behavior is obtained. Fig.~\ref{fig:ck1k2} shows
the cut of the surface $c(\kappa_{1},\kappa_{2})$ with the plane
$\kappa_{1}=-\kappa_{2}$ for $p/q=1/3$, making clearly apparent the
sub-ballistic behavior of the system for all $\kappa$. We have also
studied higher moments of order four and six. The asymptotic
behavior of these moments is consistent with the power-law behavior
of the second moment, i.e. all the moments obtained with the
Fibonacci prescription have smaller exponents than those obtained
with a periodical sequence.

At this point it is possible to ask oneself if this sub-ballistic behavior has
a quantum origin or if it also appears in the classical world. The classical
equations of motion for the Hamiltonian eq.(\ref{qkr_ham}) are given by the
standard map \cite{Reichl}%
\begin{align}
P_{n+1}  &  =P_{n}+K\sin\theta_{n}\label{standard}\\
\theta_{n+1}  &  =\theta_{n}+P_{n+1}\text{ \ }\operatorname{mod}2\pi\nonumber
\end{align}
where $n$ indicates the time step. In the phase space $\left(  \theta
_{n}\text{,}P_{n}\right)  $ of the standard map a beautiful diagram is
obtained where the Kolmogorov-Arnold-Moser (KAM) tori prevent diffusion in
angular momentum for small values of $K<K_{cri}$. The classical model
equivalent to the quantum model developed in this work is obtained when the
strength parameter $K$ takes the values $K_{1}$ and $K_{2}$ in a given
sequence in correlation with the time step. We have worked numerically with
the map eq.(\ref{standard}) in the periodic, random and aperiodic (following
the Fibonacci prescription) sequences. In the periodic case the KAM surfaces
are established, but in the random and aperiodic cases the KAM surfaces are
broken and the classical diffusion, is established. Then we can conclude that
the sub-ballistic behavior of the model is a quantum phenomenon.


Quantum resonances and DL, that were at first established in
numerical and theoretical form, have been experimentally observed
more than ten years ago in samples of cold atoms interacting with a
far-detuned standing wave of laser light
\cite{expwave,expqkr,Ammann}. This notable series of experiments
have drawn much attention because they may be establishing both the
conceptual and experimental basis of quantum computers
\cite{Santos,Benenti,Terraneo,Keating}. Experimentally only the
primary resonances of the QKR are easily observable, but recently,
Kanem $et$ $al$.\cite{Kanem}, have observed secondary resonances. On
the other hand, several experiments have been proposed
\cite{Dur,Travaglione,Sanders,Knight} to construct models of QWs.
These proposals, at first sight, have some common elements with the
experimental implementation of the QKR in the recent past. Thus, in
this theoretical and experimental frame where the QKR and the QW
have equivalent behaviors and their experimental facilities have
many elements in common, the question to be posed is: may the QKR be
considered as the fundamental model of QW? More experimental and
theoretical work is still necessary in order to answer this
question.

In summary, we have developed an aperiodic QKR model and established
a deeper equivalence between the QKR and QW. In this model we found
a new sub-ballistic behavior in the secondary resonances,
$\sigma(t)\sim t^{c}$ with $1/2<c<1$, where $c$ depends on $\kappa$,
$p$, $q$. The usual ballistic behavior of the QKR model is retained
in the primary resonances, then these resonances are robust because
they are not affected by the type of sequences of the operators
$U\left(\kappa_{1}\right)$ and $U\left(\kappa_{2}\right)$, periodic,
random or aperiodic. We explained this robustness by the
commutativity of the $U$ operators and we obtained the ballistic
diffusion coefficient analytically. Finally, we showed numerically
that the sub-ballistic behavior is a quantum interference phenomenon
that has no classical analogue.

We acknowledge the support from PEDECIBA and PDT S/C/IF/54/5 and
V.M. acknowledges the support of D.F.P.D.-ANEP.

\end{document}